\documentclass[final,1p]{elsarticle}
\usepackage{float}
\usepackage{ifpdf}
\usepackage{graphicx}
\usepackage{amssymb}
\usepackage{amsmath}
\usepackage{float}
\journal{Wave Motion}
\begin{document}

\begin{frontmatter} 

\title{On the characterization of breather and rogue wave solutions and modulation instability of a coupled generalized nonlinear Schr\"{o}dinger equations}
\author{N. Vishnu Priya}
\author{\corref{cor1}M. Senthilvelan}
\cortext[cor1]{Corresponding address: Centre for Nonlinear Dynamics, School of Physics,  Bharathidasan University, Tiruchirappalli - 620 024, Tamil Nadu, India}
\ead{velan@cnld.bdu.ac.in}
\address{Centre for Nonlinear Dynamics, School of Physics,  Bharathidasan University, \\Tiruchirappalli - 620 024, Tamil Nadu, India}





\begin{abstract}
We construct Darboux transformation of a coupled generalized nonlinear Schr\"{o}dinger (CGNLS) equations and obtain exact analytic expressions of breather and rogue wave solutions.  We also formulate the conditions for isolating these solutions.  We show that the rogue wave solution can be found only when the four wave mixing parameter becomes real.  We also investigate the modulation instability of the steady state solution of CGNLS system and demonstrate that it can occur only when the four wave mixing parameter becomes real.  Our results give an evidence for the connection between the occurrence of rogue wave solution and the modulation instability.
\end{abstract}

\begin{keyword}
Coupled generalized nonlinear Schr\"{o}dinger system \sep Modulation instability \sep Rogue waves.

\MSC{37K40 \sep 35Q51 \sep 35Q55}
\end{keyword}

\end{frontmatter}

\section{Introduction}
\label{intro}
Rogue waves (RWs) are also known as freak waves, monster waves, killer waves and extreme waves which occasionally appearing in the ocean that can reach the amplitudes more than twice the value of significant wave height \cite{Pelin}.  A well known description of RW is that it appears from nowhere and disappears without a trace \cite{Akhmediev1}.  These waves may arise from the instability of a certain class of initial conditions that tend to grow exponentially and thus have the possibility of increasing up to very high amplitudes, due to modulation instability.  It has been found that the nonlinear Schr\"{o}dinger (NLS) equation,
$iq_t+q_{xx}+2q\left\vert q \right\vert^2=0,$ where $q$ is the slowly varying pulse envelope, $x$ and $t$ are the spatial and temporal coordinates respectively, can describe many dynamical features of the RWs.  Even though they are first observed in the ocean recent investigations have shown that they also arise in optical fibers \cite{Solli}, Bose-Einstein condensates \cite{BEC}, superfluid Helium \cite{He}, capillary waves \cite{Capillary}, plasmas \cite{Plasma}, and so on.
\par In a variety of complex systems such as optical fibers, Bose-Einstein condensates, and financial systems, several amplitudes rather than a single one need to be considered. The resulting systems of coupled equations
may thus describe extreme waves with higher accuracy than the scalar NLS equation model.  Very recently, attempts have been made to construct vector rogue wave solutions of Gross-Pitaevskii equation and the Manakov system.  However, rogue waves in the coupled NLS equation with higher order effects such as four-wave mixing effects have not been studied so far.  Four-wave mixing is a basic nonlinear phenomenon having fundamental relevance and practical applications, particularly in nonlinear optics, optical processing \cite{Pepper}, phase conjugate optics \cite{Yariv}, real time holography \cite{Gerritsen} and measurement of atomic energy structures and decay rates \cite{{Bjorklund},{Yajima}}.  
\par Motivated by this, in this paper, we characterize the breather and RW solutions of a physically important system, namely coupled generalized nonlinear Schr\"{o}dinger (CGNLS) system \cite{{Wang},{Penj}},
\begin{subequations}
\begin{align}
ip_t+p_{xx}+2(a\vert p\vert ^{2}+c\vert q\vert ^{2}+bpq^{*}+b^{*}qp^{*})p &= 0, \\
iq_t+q_{xx}+2(a\vert p\vert ^{2}+c\vert q\vert ^{2}+bpq^{*}+b^{*}qp^{*})q &= 0,
\end{align}
\label{pct1}
\end{subequations}
where $p$ and $q$ are slowly varying pulse envelopes and $a$ and $c$ are real constants.  Here $b$ is a complex constant and $*$ denotes complex conjugation.  Physically $a$ and $c$ terms describe the self phase modulation and cross phase modulation effects respectively.  The $b$ and $b^*$ terms describe the four wave mixing effects.  When $a=c$ and $b=0$ the above equations reduce to the Manakov system \cite{Manakov}.  When $a=-c$ and $b=0$ they reduce to the mixed coupled nonlinear Schr\"{o}dinger equation \cite{viji}.  
\par Although a connection between RW solution and modulation instability of NLS equation has been analyzed explicitly,  to our knowledge, the connection between RWs in the coupled NLS equation with higher order effects such as four-wave mixing effects and modulation instability has not been studied so far.  In this paper, besides characterizing breather and RW solutions of (\ref{pct1}), we also intend to investigate the effect of four-wave mixing in modulation instability process.    
\par We derive breather and RW solutions of CGNLS system (\ref{pct1}) using Darboux transformation (DT).  DT is one of the powerful methods to construct a class of solutions including, N-soliton solutions (bright and dark types), breathers and RW solutions for the nonlinear integrable evolution equations.  The DT retains the form of linear eigenvalue problem under a specific transformation and generate new solutions from the seed solution.  To obtain breather and RW solutions we choose plane wave solution as the seed solution.  Since the Lax pair matrices of (\ref{pct1}) is of order $3\times 3$, while integrating the Lax pair equations we obtain a cubic polynomial equation. The cubic polynomial may have (i) three different roots, (ii) one single and one double root and (iii) three equal roots.  Since the general solution depends on the roots of the characteristic equation, the dynamics comes out from the above three cases also differ from each other.  In the first and second cases we obtain breather and RW solution respectively for the Eq. (\ref{pct1}).  In both the cases we obtain the respective solutions by considering the four wave mixing parameter $b$ as a real constant.  In the third case we obtain a trivial solution only.  We then examine the modulation instability of the CGNLS system and demonstrate that it can occur when the imaginary part of the four wave mixing parameter $b$ should be zero.  From these two results we conclude that the RW can occur only in the regime of modulation instability.  Very recently breather and RW solutions have also been obtained for the Manakov system \cite{{Degasperis},{josab}}.  Since CGNLS system is a generalization of Manakov system we also compare the results obtained in this paper with the Manakov equation and point out the differences in the breather and RW solutions that present in these two systems.
\par The plan of the paper is as follows.  To begin with, in the following section, we briefly summarize the results obtained for the equation (\ref{pct1}).  In section 3, we construct DT of CGNLS system (\ref{pct1}).  In section 4, we derive general breather and RW solutions of CGNLS system.  In section 5, we investigate modulation instability of CGNLS system and point out the connection between RW and modulation instability.  Section 6 is devoted to conclude our work.

\section{Summary of work done on (\ref{pct1})}
In this section, we briefly summarize the results obtained on the CGNLS system during the past few years.  The CGNLS system (\ref{pct1}) was first introduced as an completely integrable equation by Wang et al \cite{Wang}.  With the help of Riemann-Hilbert method the authors have constructed N-bright soliton solutions for (\ref{pct1}) with a condition on system parameters $a$, $b$ and $c$.  The collision dynamics between two bright solitons and the existence of infinite number of conservation laws are also shown in the same reference \cite{Wang}.  In a subsequent study L\"{u} and Penj have carried out Painlev$\acute{e}$ singularity structure analysis to this equation and shown that this system passes P-test for arbitrary values of system parameters $a$, $b$ and $c$ \cite{Penj}.  In one of our recent studies, we have constructed dark-dark soliton and general breather solutions for the Eq. (\ref{pct1}) through Hirota's bilinearization method \cite{vishnu1}.  By appropriately restricting the parameters which appear in the phase factors in the general breather expression we have derived Akhmediev breather, Ma soliton and rogue wave solutions \cite{vishnu1}.  We then considered the reverse problem of constructing Akhmediev breather, Ma soliton and general breather (GB) expression from the RW solution.  To derive various breather profiles from RW solution we considered the later in a factorized form and generalized this factorized form in an imbricate series expression with certain unknown parameters in it.  By substituting this imbricate series in Eq. (\ref{pct1}) and solving the resultant equations we fixed the unknown parameters.  Substituting these parameters back in the imbricate series and rewriting it in a compact form we have obtained the desired breather solutions.  With three different forms of the imbricate series we have derived the Akhmediev breather, Ma soliton and general breather solutions from the RW solution \cite{vishnu1}.  We have also studied in-detail how these solution profiles change with respect to four wave mixing terms.  We have demonstrated that when we increase the value of the real value of four wave mixing parameter not only the number of peaks increases in the GB profile but also the direction of the underlying profile gets altered.  When we increase the  value of $Re$ $b$ in the AB and MS profiles only the number of peaks increases in them.  In the case of RW solution the profile becomes thinner and thinner when we increase the value of $Re$ $b$ \cite{vishnu1}.  We observed that the value of $Im$ $b$ does not change either the number of pulses or the direction of any of these profiles.  
\par In a follow-up work, we have constructed Nth order RW solution to CGNLS system (\ref{pct1}) using generalized Darboux transformation (GDT) method \cite{vishnu2}.  It is very difficult to construct higher order rogue wave solutions using classical DT method since higher order rogue wave solutions do contain only one critical eigenvalue.  It is well known that in the conventional N-fold DT one has N-distinct eigenvalues.  Through GDT one can overcome this problem and construct higher order rogue waves with single eigenvalue.  Using GDT we have derived a recursive formula and presented determinant representations for Nth order RW solution of this system.  Using these representations we have displayed second order RW solution with two free parameters and the third order RW solution with four free parameters \cite{vishnu2}.  We have also analyzed the second and third order RW solutions by varying these free parameters and shown that one can obtain certain interesting structures.  In the case of second order RW, we have shown that these RWs exhibit a triplet pattern and in the case of third order RW solution we have visualized a triangular and hexagonal structures depending upon the restriction on the free parameters \cite{vishnu2}.  In continuation of earlier studies, in this work, we aim to investigate the interaction behaviors in nonlinear localized modes such as solitons, breather and rogue wave in the CGNLS system.

\section{Darboux transformation (DT) of CGNLS system}
We rewrite the CGNLS system (\ref{pct1}) in a matrix form as
\begin{eqnarray}
P_t-\frac{1}{2}\tilde{P}_x^2+P^2\tilde{P}=0,
\label{eqn}
\end{eqnarray}
with
\begin{eqnarray}
P=\begin{pmatrix}0&0&p\\0&0&q\\r_1&r_2&0\end{pmatrix}\; {\text{and}}\;J=\begin{pmatrix}i&0&0\\0&i&0\\0&0&-i\end{pmatrix},
\label{PJ}
\end{eqnarray}
where $\tilde{P}=[J,P]$, $r_1=-(ap^*+bq^*)$ and $r_2=-(b^*p^*+cq^*)$.  One can easily check Eq. (\ref{eqn}) leads to the CGNLS system (\ref{pct1}).  We can write the associated linear equation as
\begin{eqnarray}
\left(\partial_x-P-\lambda J\right)\Psi=0,\nonumber\\
\left(\partial_t-\frac{1}{2}(P\tilde{P}+\tilde{P}_x)+2\lambda P+2\lambda^2J\right)\Psi=0,
\label{laxpair}
\end{eqnarray}
where $\lambda$ is a complex spectral parameter and $\Psi(x,t,\lambda)$ is a three component vector.  If the matrix $P$ satisfies (\ref{eqn}), one can find a nonzero solution $\Psi$ by integrating the Lax pair equation (\ref{laxpair}) \cite{Park}.
\par We construct DT of (\ref{pct1}) by following steps given below.\\
{\bf Step 1:} We choose plane wave solution as the seed solution for $P$, that is $P=P^{(0)}$ with $p=p^{(0)}$, $q=q^{(0)}$.  \\
{\bf Step 2:} We integrate the Lax pair equations (\ref{laxpair}) to obtain the linear function $\Psi(x,t,\lambda)$.  \\
{\bf Step 3:} We then consider the complex spectral parameter as $\lambda=\lambda_1$ and evaluate the linear function $\Psi(x,t,\lambda)$ at this value, that is $\Psi_1\equiv \Psi(x,t,\lambda=\lambda_1)$.  We introduce two orthogonal auxiliary vectors, $V$ and $W$ to $\Psi_1$, such that
\begin{eqnarray}
<\Psi|V>\equiv\Psi^\dagger V=0,\; <\Psi|W>\equiv\Psi^\dagger W=0.
\label{vw}
\end{eqnarray}
{\bf Step 4:} Using these vectors we define a Darboux matrix $D$ which is of the form,
\begin{eqnarray}
D=[\Psi_1, V, W].
\end{eqnarray} \\
{\bf Step 5:} Finally, we obtain first iterated $DT$ solution of CGNLS system from,
\begin{eqnarray}
p[1]=p+2i\frac{|N_1|}{|D|},\;\;q[1]=q+2i\frac{|N_2|}{|D|},
\label{pc9}
\end{eqnarray}
where $N_i$, $i=1,2$, are the matrices obtained by replacing the third row of the matrix $D$ with the $i$th row of  the matrix $D$ in which the three elements will be multiplied with $\lambda_1$, $\lambda_1^*$ and $\lambda_1^*$ respectively.
\section{Breather and rogue wave solutions of (\ref{pct1})}
We construct breather and RW solutions of (\ref{pct1}) by following the procedure outlined in previous section.  To begin with, we consider the seed solution be of the form
\begin{eqnarray}
p[0]=a_1e^{i(k_1x+\omega_1t)}, \;\; q[0]=a_2e^{i(k_2x+\omega_2t)},
\end{eqnarray}
where $a_1$, $a_2$, $k_1$, $k_2$, $\omega_1$ and $\omega_2$ are real constants.  Substituting the above forms in Eq. (\ref{pct1}), we get
\begin{eqnarray}
-a_1\omega_1e^{i(k_1x+\omega_1t)}-a_1k_1^2e^{i(k_1x+\omega_1t)}+2(aa_1^2+ca_2^2+ba_1a_2e^{i((k_1-k_2)x+(\omega_1-\omega_2)t)}\nonumber\\+b^*a_1a_2e^{i((k_2-k_1)x+(\omega_2-\omega_1)t)})a_1e^{i(k_1x+\omega_1t)}=0,\nonumber\\
-a_2\omega_2e^{i(k_2x+\omega_2t)}-a_2k_2^2e^{i(k_2x+\omega_2t)}+2(aa_1^2+ca_2^2+ba_1a_2e^{i((k_1-k_2)x+(\omega_1-\omega_2)t)}\nonumber\\+b^*a_1a_2e^{i((k_2-k_1)x+(\omega_2-\omega_1)t)})a_2e^{i(k_2x+\omega_2t)}=0.
\end{eqnarray}
The presence of four wave mixing parameter $b$ and $b^*$ hinders us to obtain a dispersion relation.  To overcome this obstacle we restrict the parameters to be $k_1=k_2$ and $\omega_1=\omega_2$, so that we can get a consistent dispersion relation of the form
\begin{eqnarray}
\omega_1=-k_1^2+2h, \; h=(aa_1^2+ca_2^2+2b_Ra_1a_2).
\label{dispersion} 
\end{eqnarray}
 With this restriction the seed solution become,
\begin{eqnarray}
p[0]=a_1e^{ik_1x+\omega_1t}\;\; q[0]=a_2e^{ik_1x+\omega_1t}.\label{seed}
\end{eqnarray}
\par In the second step we integrate the Lax pair equations (\ref{laxpair}).  Considering $\Psi$ as $\Psi=(\psi(x,t), \phi(x,t), \varphi(x,t))$, Eq. (\ref{laxpair}) yields the following six coupled equations, namely 
\begin{align}
\psi_{1x} = & i\lambda_1\psi_1+p\varphi_1,\;\;
\phi_{1x} = i\lambda_1\phi_1+q\varphi_1,\;\;
\varphi_{1x} = r_1\psi_1+r_2\phi_1-i\lambda \varphi_1,\nonumber\\
\psi_{1t} = & (-2i\lambda_1^2-ipr_1)\psi_1-ipr_2\phi_1+(ip_x-2p\lambda_1)\varphi_1,\nonumber\\
\phi_{1t} = & (-2i\lambda_1^2-iqr_2)\phi_1-iqr_1\psi_1+(iq_x-2q\lambda_1)\varphi_1,\nonumber\\
\varphi_{1t} = & (-ir_{1x}-2\lambda_1r_1)\psi_1+(-ir_{2x}-2\lambda_1r_2)\phi_1+(ipr_1+iqr_2+2i\lambda_1^2)\varphi_1.
\label{l6}
\end{align}
\par Substituting the seed solution (\ref{seed}) into (\ref{l6}) and integrating the resultant set of differential equations (\ref{l6}) we can get the eigenfunction $\Psi=(\psi_1,\phi_1,\varphi_1)$.  Integrating the system of differential equations (\ref{l6}) we come across the following characteristic polynomial equation, namely
\begin{eqnarray}
&&\tau^3-i(k_1+\lambda)\tau^2-(a_1(-aa_1-bs_2)-\lambda^2+2k_1\lambda+a_2(-b^*a_1-ca_2))\tau\nonumber\\&&+i\lambda(a_1(-aa_1-bs_2)-\lambda^2+k_1\lambda+a_2(-b^*a_1-ca_2))=0.
\label{spathird}
\end{eqnarray}
The cubic polynomial (\ref{spathird}) may have (i) three different roots, (ii) one single and one double root and (iii) three equal roots.  The general solution $\Psi$ depends on the roots of the Eq. (\ref{spathird}).  We integrate the set of Eqs. (\ref{l6}) by considering all three possibilities and report the outcome.  

\subsection{Case 1}
In the first case, that is all the three roots of the characteristic equation are different, we obtain the following expressions for $\psi$, $\phi$ and $\varphi$, namely
\begin{eqnarray}
\psi(x,t)&=&a_1\left(c_1 e^A+c_2 e^B+c_3e^{i\lambda x-2i\lambda^2t}\right),\nonumber\\
\phi(x,t)&=&a_2\left(c_1 e^A+c_2 e^B-\frac{a_1(aa_1+ba_2)}{a_2(ca_2+b^*a_1)}c_3e^{i\lambda x-2i\lambda^2t}\right),\nonumber\\
\varphi(x,t)&=&e^{-i(k_1x+\omega_1t)}\left(c_1L_+e^A+c_2L_-e^B\right)\label{case1},
\end{eqnarray} 
where $A=\frac{1}{2}(ik_1-\sqrt{s})x+((\frac{1}{2}k_1+\lambda)\sqrt{s}-\frac{i}{2}k_1^2+ih)t$, $B=\frac{1}{2}(ik_1+\sqrt{s})x+(-(\frac{1}{2}k_1+\lambda)\sqrt{s}-\frac{i}{2}k_1^2+ih)t$, $s=-k_1^2-4(h-k_1\lambda+\lambda^2)$, $L_{\pm}=\frac{1}{2}(ik_1\pm\sqrt{s})-i\lambda$ and $\lambda=f+ig$, where $a_1$, $a_2$, $k_1$, $f$ and $g$ are real parameters and $c_1$, $c_2$ and $c_3$ are integration constants.  
\par To proceed further we choose two auxiliary vectors which are of the form (Step 3), 
\begin{eqnarray}
V=\left(\varphi_1^*,0,-\psi_1^*\right)^T,\;\; W=\left(0, \varphi_1^*, -\phi_1^*\right)^T.
\label{vw1}
\end{eqnarray}
These auxiliary vectors are consistent with the solutions (\ref{case1}) provided the following conditions are satisfied, namely
\begin{eqnarray}
(a-1)+b\frac{a_2}{a_1}=0,\;\;
b^*+(c-1)\frac{a_2}{a_1}=0,\label{con2}\;\;c_3=0.
\label{con}
\end{eqnarray}
\par Now we can define a Darboux matrix $D$ of the form (Step 4)
\begin{eqnarray}
D=\left(\begin{array}{ccc}\psi_1&\varphi_1^*&0\\\phi_1&0&\varphi_1^*\\\varphi_1&-\psi_1^*&-\phi_1^*\end{array}\right),
\label{matrixD}
\end{eqnarray}
so that the matrices $N_i$, $i=1,2$, turn out to be
\begin{eqnarray}
N_1=\left(\begin{array}{ccc}\psi_1&\varphi_1^*&0\\\phi_1&0&\varphi_1^*\\\lambda_1\psi_1&\lambda_1^*\varphi_1^*&0\end{array}\right)\;{\text{and}}\;N_2=\left(\begin{array}{ccc}\psi_1&\varphi_1^*&0\\\phi_1&0&\varphi_1^*\\\lambda_1\phi_1&0&\lambda_1^*\varphi_1^*\end{array}\right).
\label{matrixN}
\end{eqnarray}
Substituting the matrices $N_i$ and $D$ in (\ref{pc9}) and evaluating the determinants we obtain the exact form of first iterated DT solution of CGNLS system as (Step 5)
\begin{eqnarray}
p[1]&=&p+2i\frac{(\lambda_1-\lambda_1^*)\psi_1\varphi_1^*}{|\psi_1|^2+|\phi_1|^2+|\varphi_1|^2},\nonumber\\
q[1]&=&q+2i\frac{(\lambda_1-\lambda_1^*)\phi_1\varphi_1^*}{|\psi_1|^2+|\phi_1|^2+|\varphi_1|^2}.
\label{pct9}
\end{eqnarray}
\par Considering the expressions $\psi$, $\phi$ and $\varphi$ given in (\ref{case1}) with $c_3=0$ and substituting them in (\ref{pct9}) and simplifying the resultant expressions we arrive at the following general breather solution of CGNLS system, that is 
\begin{eqnarray}
p&=&a_1e^{i(k_1x+\omega_1t)}\nonumber\\&&\times\left(1+\frac{2i(\lambda-\lambda^*)(L_+^*c_1^2e^{A+A^*}+L_-^*c_1c_2e^{A+B^*}+L_+^*c_1c_2e^{A^*+B}+L_-^*c_2^2e^{B+B^*})}{G_1c_1^2e^{A+A^*}+G_2c_1c_2e^{A+B^*}+G_3c_1c_2e^{A^*+B}+G_4c_2^2e^{B+B^*}}\right), \nonumber\\
q&=&a_2e^{i(k_1x+\omega_1t)}\nonumber\\&&\times\left(1+\frac{2i(\lambda-\lambda^*)(L_+^*c_1^2e^{A+A^*}+L_-^*c_1c_2e^{A+B^*}+L_+^*c_1c_2e^{A^*+B}+L_-^*c_2^2e^{B+B^*})}{G_1c_1^2e^{A+A^*}+G_2c_1c_2e^{A+B^*}+G_3c_1c_2e^{A^*+B}+G_4c_2^2e^{B+B^*}}\right),\nonumber\\
\label{case1sol}
\end{eqnarray}
with
\begin{eqnarray}
G_1&=&a_1^2+a_2^2+L_+L_+^*, \; G_2=a_1^2+a_2^2+L_+L_-^*, \nonumber\\
G_3&=&a_1^2+a_2^2+L_-L_+^*,  \; G_4=a_1^2+a_2^2+L_-L_-^*.
\end{eqnarray}
\par The general breather solution (\ref{case1sol}) satisfies the CGNLS equations with the conditions given in (\ref{con}).  The solution (\ref{case1sol}) given above, is periodic both in space and time directions as shown in Fig.1.  We mention here that the components $p$ and $q$ are linearly dependent.  This is because of the fact that to obtain a consistent dispersion relation (vide Eq. (\ref{dispersion})) we have forced to restrict the parameters, $k_1=k_2$ and $\omega_1=\omega_2$.  In other words the seed solution $p[0]$ and $q[0]$ which we have picked up differ only in amplitude.  The solutions which come out from this seed solution subsequently differ only in amplitude.  As far as the Manakov system \cite{josab} is concerned while determining the dispersion relation  we do not come across this constraint and so we can choose different plane wave solutions for each component.  Because of this freedom, in the Manakov system, in the three different roots case, one can explore different interaction structures between nonlinear modes such as dark soliton-Akhmediev breather interactions, Akhmediev breather-Akhmediev breather interactions and so on. 
\begin{figure}
\includegraphics[width=1\linewidth]{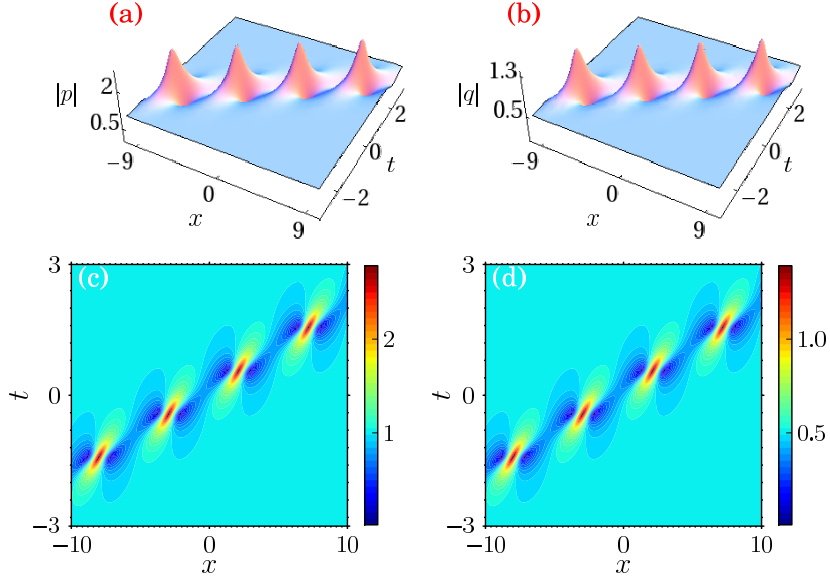}
\caption{(a) General breather profile of $p$ component (b) General breather profile of $q$ component for the parameter values $f=0.5$, $g=1$, $a_1=1$, $a_2=0.5$, $b_R=2$, $c_1=0.5$, $c_2=0.6$.  (c) and (d) are corresponding contour plots for $p$ and $q$ components respectively.}
\end{figure}       
\subsection{Case 2}
In the previous subsection we examined the three different roots case.  In this sub-section, we investigate the case two equal roots and one single root in the characteristic polynomial equation.  In this case we have a restriction between the spectral parameter $\lambda$ and system parameters $a$, $c$ and $b$, namely $\lambda=\frac{k_1}{2}\pm i\sqrt{h}.$  Solving the system of Eqs. (\ref{l6}) with this restriction we arrive at the following expressions for $\psi$, $\phi$ and $\varphi$, namely
\begin{eqnarray}
\psi(x,t)&=&c_3e^C+\frac{a_1}{2(\sqrt{h}-ik_1)}e^D[2c_1\sqrt{h}+2c_2\sqrt{h}t-2ik_1(c_1+c_2t)+ic_2x],\nonumber\\
\phi(x,t)&=&-\frac{aa_1+ba_2}{b^*a_1+ca_2}c_3e^C+\frac{a_2}{2(\sqrt{h}-ik_1)}e^D[2c_1\sqrt{h}+2c_2\sqrt{h}t-2ik_1(c_1+c_2t)+ic_2x],\nonumber\\
\varphi(x,t)&=&\frac{e^{-D}}{2(\sqrt{h}-ik_1)}[2c_1(h-ik_1\sqrt{h})+c_2(i+2ht-2i\sqrt{h}t+i\sqrt{h}x)],
\label{case2}
\end{eqnarray}
where $C=(2k_1\sqrt{h}-\frac{i}{2}(k_1^2-4h))t+(\frac{ik_1}{2}-\sqrt{h})x$, $D=\frac{i}{2}(k_1x+(2h-k_1^2)t)$ and $h$ is defined in (\ref{dispersion}).  
\par Once the exact form of $\Psi_1=(\psi_1,\phi_1,\varphi_1)$ is determined we then move on to construct two auxiliary vectors $V$ and $W$ (Step 3).  We find that the auxiliary vectors given in (\ref{vw1}) are consistent with the solutions (\ref{case2}) provided they satisfy the conditions given in (\ref{con}).  Since the auxiliary vectors are same as given in (\ref{vw1}) the matrices $D$, $N_i$, $i=1,2$ and the form of first iterated solution are all coincide with the expressions given in (\ref{matrixD}), (\ref{matrixN}) and (\ref{pct9}) respectively (Steps 4 and 5).   
\par Substituting the expressions given in (\ref{case2}) with $c_3=0$ in (\ref{pct9}) we can get a RW solution of the form, 
\begin{eqnarray}
p=a_1e^{i(k_1x+\omega_1t)}\left(1-4\sqrt{h}\frac{N}{D}\right), \;\; q=a_2e^{i(k_1x+\omega_1t)}\left(1-4\sqrt{h}\frac{N}{D}\right),
\label{case2sol}
\end{eqnarray}
where
\begin{eqnarray}
N&=&2c_2(\sqrt{h}-ik_1)(-ic_3+2c_2(h+ik_1\sqrt{h}))+(2c_2c_3(k_1^2+h)\sqrt{h}+2c_3^2(-k_1-i\sqrt{h}))t\nonumber\\&&+4c_3^2\sqrt{h}(k_1^2+h)t^2+(c_3^2-4c_2c_3k_1\sqrt{h})x-4c_3^2k_1\sqrt{h}xt+c_3^2\sqrt{h}x^2,\nonumber\\
D&=&(a_1^2+a_2^2)(4(k_1^2+h)(c_2+c_3t)^2-4c_3k_1(c_2+c_3t)x+c_3^2x^2)+c_3^2-4c_2c_3k_1\sqrt{h}\nonumber\\&&+4c_2^2(k_1^2+h)h+4c_3(2c_2h+2c_2h^2-c_3k_1\sqrt{h})t+4c_3^2h(k_1^2+h)t^2\nonumber\\&&+2c_3(c_3\sqrt{h}-2c_2k_1h)x-4c_3^2k_1hxt+c_3^2hx^2,
\end{eqnarray}
where $\omega$ and $h$ are defined in (\ref{dispersion}).  It should be noted that the parameter $h$ and the RW solution (\ref{case2sol}) contains only the real part of four wave mixing parameter $b_R$.  The RW solution is localized both in spatial and time directions as shown in Fig. 2.  

\par Here also one may observe that the components are linearly dependent.  They differ in amplitude only.  This is again because of the restriction imposed on the parameters $k_i'$s and $\omega_i'$s, $i=1,2$, in the seed solution.   As far as the Manakov system \cite{josab} is concerned, in a double root and a single root case, no restriction on wave numbers $k_1$ and $k_2$ is imposed and because of this freedom, one can explore certain interaction structures such as bright soliton - RW interaction, dark soliton-RW interaction, Akhmediev breather-RW interactions and dark rogue wave structure \cite{josab}.  The presence of four wave mixing effects in the CGNLS system destroy these interaction behaviors.    

\begin{figure}
\includegraphics[width=1\linewidth]{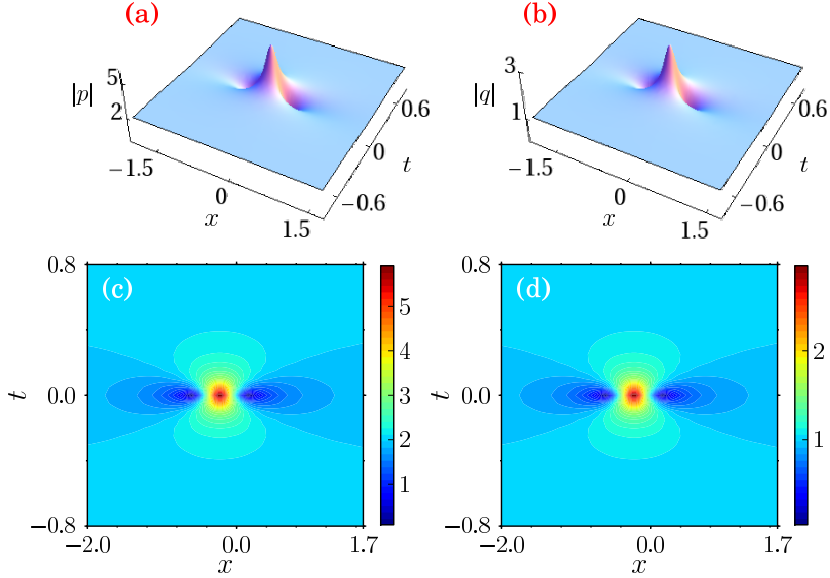}
\caption{(a) First order RW profile of the $p$ component, (b) First order RW profile of the $q$ component for the parameter values $a_1=2$, $a_2=1$, $b_R=1$, $c_2=0.5$, $c_3=1$.  (c) and (d) are corresponding contour plots for $p$ and $q$ components respectively.}
\end{figure}         
\subsection{All the three roots are equal}
Finally, we consider the case all the three roots are equal.  The underlying roots can be determined as follows.  Let us consider a cubic polynomial Eq. (\ref{spathird}) be in the form
\begin{eqnarray}
x^3+\alpha x^2+\beta x+\gamma=0.
\label{c1}
\end{eqnarray}
Eq. (\ref{c1}) has three equal roots provided that the coefficients $\alpha$, $\beta$ and $\gamma$ satisfy the following two conditions, namely
\begin{eqnarray}
\alpha^2-3\beta=0, \;\; 2\alpha^3-9\alpha\gamma+27\gamma=0. \label{c2}
\end{eqnarray}
where
\begin{eqnarray}
\alpha=-i(k_1+\lambda),\;\; \beta=-(a_1(-aa_1-ba_2)-\lambda^2+2k_1\lambda+a_2(-b^*a_1-ca_2))\nonumber\\
\gamma=i\lambda(a_1(-aa_1-ba_2)-\lambda^2+k_1\lambda+a_2(-b^*a_1-ca_2)).
\label{alpha}
\end{eqnarray} 
Substituting (\ref{alpha}) in the first condition in (\ref{c2}) and rewriting the resultant expression for $\lambda$, we find
\begin{eqnarray}
\lambda=\frac{1}{2}(k_1\pm i\sqrt{3}\sqrt{h}).\label{lambda}
\end{eqnarray}
Substituting this expression in the second condition in (\ref{c2}) we end up at $h=0$.  With this restriction the expression given for $\lambda$ (\ref{lambda}) becomes real, that is $\lambda=\frac{k_1}{2}$.  If $\lambda$ becomes real, the DT formula (\ref{pct9}) provides only a trivial solution.  So we conclude that the third case does not provide any new solution.  
\par In the Manakov system \cite{josab}, in the triple root case, due to the conditions given in (\ref{c2}), we will a get restriction on the wave numbers in the form
\begin{eqnarray}
|k_1-k_2|=\sqrt{g_1}s_1,\label{manakov}
\end{eqnarray}
where self phase modulation parameter $g_1$ is related to the parameter $a$ given in this paper (vide Eq. (\ref{pct1})) and the amplitude of plane wave $s_1$ is related with $a_1$  (vide Eq. (10)) and $k_1$ and $k_2$ are the wave numbers of the plane waves.  The condition (\ref{manakov}) was also given in Ref. \cite {josab}.  As far as the CGNLS system is concerned we have restricted the plane wave backgrounds with $k_1=k_2$.  The left hand side in Eq. (\ref{manakov}) becomes zero with this restriction and choose $s_1=0$ on the right hand side to make the condition (\ref{manakov}) consistent.  The restriction $s_1=0$ indicates a trivial solution.  In this sense  our result is consistent with the one reported in \cite{josab}.

\section{Modulation instability and the criterion for existence of RW}
The initial exponential growth of spectral sidebands and subsequent nonlinear evolution have been discussed theoretically in terms of breathers in optical experiments through fiber and photonic crystal wave guides.  In the following, we derive a connection between MI and the criterion for the existence of a RW in Eq. (\ref{pct1}). 
\par To study the MI, we first consider a steady state solution of Eq. (\ref{pct1}),
\begin{eqnarray}
p=a_1e^{i\omega t},\;\;q=a_2e^{i\omega t},
\label{ststsol}
\end{eqnarray}
with $\omega=2(aa_1^2+ca_2^2+(b+b^*)a_1a_2)$.  We then check whether this steady state solution is stable against small perturbations or not.  To do this, we perturb the steady state solution in a way such that,
\begin{eqnarray}
p=(a_1+U_1)e^{i\omega t},\;\;q=(a_2+U_2)e^{i\omega t},
\label{pertsol}
\end{eqnarray}
where $U_j(x,t)$, $j=1,2,$ are weak perturbations.  Substituting Eq. (\ref{pertsol}) in (\ref{pct1}) and linearizing the equations in $U_1$ and $U_2$, we find
\begin{eqnarray}
iU_{1t}+U_{1xx}+2[aa_1^2(U_1+U_1^*)+ca_1a_2(U_2+U_2^*)+b(a_1^2U_2^*+a_1a_2U_1)\nonumber\\+b^*(a_1^2U_2+a_1a_2U_1^*)]=0,\nonumber\\
iU_{2t}+U_{2xx}+2[aa_1a_2(U_1+U_1^*)+ca_2^2(U_2+U_2^*)+b(a_1a_2U_2^*+a_2^2U_1)\nonumber\\+b^*(a_1a_2U_2+a_2^2U_1^*)]=0.  
\label{perteqn}
\end{eqnarray} 
To solve these equations we assume a general solution of the form
\begin{eqnarray}
U_j=u_j\cos(Kt-\Omega x)+iv_j\sin(Kt-\Omega x), \;\; j=1,2,
\label{pertgensol}
\end{eqnarray}
where $K$ is the wave number and $\Omega$ is the frequency of perturbation.  Substituting the assumed solutions  into Eqs. (\ref{perteqn}) and splitting the resultant equations into real and imaginary parts, we obtain the following set of four algebraic equations, namely
\begin{eqnarray}
(-\Omega^2+4aa_1^2+4b_Ra_1a_2)u_1+(4ca_1a_2+4b_Ra_2^2)u_2-Kv_1=0,\nonumber\\
-Ku_1-\Omega^2v_1=0,\nonumber\\
(4aa_1a_2+4b_Ra_2^2)u_1+(-\Omega^2+4aa_2^2+4b_Ra_1a_2)u_2-Kv_2=0,\nonumber\\
-Ku_2-\Omega^2v_2=0.
\label{mi1}
\end{eqnarray}
\begin{figure}
\includegraphics[width=0.8\linewidth]{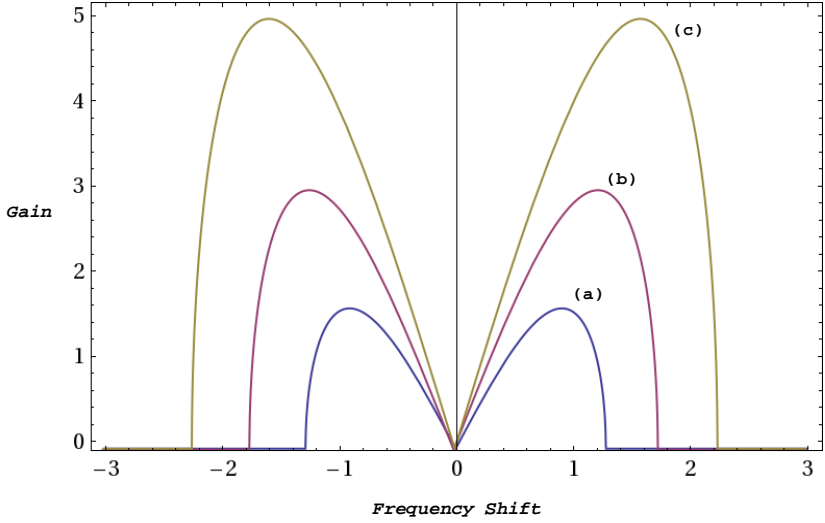}
\caption{Gain spectra of modulation instability for the parameter values $a=1$, $c=1$, $b_R=0.5$ and for the different power levels (a) $a_1=0.5$, $a_2=0.6$, (b) $a_1=1$, $a_2=1.5$, (c) $a_1=2$, $a_2=2.5$.}
\end{figure}
We derived the above set of equations with the restriction that $b$ should be a real constant.
The system of Eqs. (\ref{mi1}) have nontrivial solutions only if the determinant of the coefficient matrix
vanishes.  In our case we find that the determinant vanishes provided that the following condition is fulfilled, that is
\begin{eqnarray}
K^4+(4aa_1^2+4ca_2^2+8b_Ra_1a_2-2\Omega^2)\Omega^2K^2-(4aa_1^2+4ca_2^2+8b_Ra_1a_2-\Omega^2)\Omega^6=0.
\label{Keqn}
\end{eqnarray}
Solving Eq. (\ref{Keqn}) we obtain the following dispersion relation 
\begin{eqnarray}
K=\pm\Omega^2;\;\; K=\pm\sqrt{(-4aa_1^2-4ca_2^2-8b_Ra_1a_2+\Omega^2)\Omega^2}.
\label{disrel}
\end{eqnarray}
The dispersion relation (\ref{disrel}) clearly shows that the stability of the steady state depends critically on whether light experiences self-focusing or self-defocusing effects inside the fibre.  In the case of self-defocusing effect ($a,c,b_R<0$) $K$ is real for all values of $\Omega$, and the steady state is stable against small perturbations.  By contrast, in the case of self-focusing effect ($a,c,b_R>0$), $K$ becomes imaginary for $\Omega<\Omega_c$, where $\Omega_c^2=4aa_1^2+4ca_2^2+8b_Ra_1a_2$, and the perturbations $U_j(x,t)$, $j=1,2,$ grow exponentially with $t$.  This instability is referred to as the MI since it leads to a spontaneous modulation of the steady state.  Let us consider the gain spectrum of MI.  The power gain at any frequency $\Omega$ is obtained from Eq. (\ref{disrel}) and is given by
\begin{eqnarray}
g(\Omega)=2\;Im(K)=|\Omega|(\Omega^2-(4aa_1^2+4ca_2^2+8b_Ra_1a_2))^{1/2},
\end{eqnarray} 
where $g(\Omega)$ represents the gain at the frequency $|\omega|<(4aa_1^2+4ca_2^2+8b_Ra_1a_2)$.  Fig. 3 shows the gain spectra at three power levels.   
\par MI occurs in (\ref{pct1}) only when the four wave mixing parameter $b$ becomes real.  Interestingly we have also observed that the RW solution does admit only the real part of four wave mixing parameter (vide Eq. (\ref{case2sol})).  From these results we conclude that the self-focusing CGNLS Eq. (\ref{pct1}) admits MI when the four wave mixing parameter becomes real in which the RW mode can also occur.  This in turn gives an evidence for the connection between the occurrence of the RW and MI in (\ref{pct1}).  
 
\section{Conclusion}
In this work, we have studied breather and RW solutions of a CGNLS equations.  To construct these solutions we have employed DT method.  We have chosen plane wave solution as the seed solution while integrating the the Lax pair equations and obtained the eigenfunctions.  While solving the system of ODEs we came across a third order characteristic polynomial equation whose roots determine the nature of the solutions.  When the roots of polynomial equation are different we have obtained the general breather solution.  In the case of two roots are equal and the third root is different we have obtained the RW solution.  In the third case, that is all three roots are equal, we end up at a trivial solution.  We have also compared our results with the Manakov equation.  In the case of Manakov system one can choose two different plane wave solutions as the seed solution.  Due to this advantage one can get different interaction properties between nonlinear waves such as bright soliton, dark soliton, Akhmediev breather, bright RW, dark RW and double RW structures.  But in the case of CGNLS system we had a restriction on plane wave seed solution, which in turn narrows down certain localized structures.  As we have demonstrated in this paper one can obtain only the general breather and RW solutions for the CGNLS system.  Currently, we are investigating the possibility of obtaining more general solution for this system.  In addition to the above, we have analyzed the MI of the CGNLS system and found that it occurs only when the four wave mixing parameter becomes real.  RW solution also exists only for the real value of this parameter.  This coincidence establishes the connection between the occurrence of RW solution and the MI in this model.  
\section*{Acknowledgements}                      
NVP wishes to thank the University Grants Commission (UGC-RFSMS), Government of India, for providing a Research Fellowship. The work of MS forms part of a research project sponsored by National Board for Higher Mathematics (NBHM), Government of India.

\end{document}